\documentclass[twocolumn]{ceurart}

\usepackage{natbib}
\usepackage{graphicx}
\usepackage{enumitem}

\begin{document}

\copyrightyear{2024}
\copyrightclause{Copyright for this paper by its authors.
  Use permitted under Creative Commons License Attribution 4.0
  International (CC BY 4.0).}

\conference{PatentSemTech'24: 5th Workshop on Patent Text Mining and Semantic Technologies, July 18, 2024, Washington D.C., USA.}

\title{ClaimCompare: A Data Pipeline for Evaluation of Novelty Destroying Patent Pairs}

\author{Arav Parikh}[%
orcid=0000-0002-0877-7063,
email=arav.parikh@uconn.edu
]
\author{Shiri Dori-Hacohen}[%
orcid=0000-0002-0877-7063,
email=shiridh@uconn.edu
]
\address{University of Connecticut, School of Computing, Reducing Information Ecosystem Threats (RIET) Lab}

\begin{abstract}
  A fundamental step in the patent application process is the determination of whether there exist prior patents that are \textbf{novelty destroying}. This step is routinely performed by both applicants and examiners, in order to assess the novelty of proposed inventions among the millions of applications filed annually. However, conducting this search is time and labor-intensive, as searchers must navigate complex legal and technical jargon while covering a large amount of legal claims. Automated approaches using information retrieval and machine learning approaches to detect novelty destroying patents present a promising avenue to streamline this process, yet research focusing on this space remains limited. In this paper, we introduce a novel data pipeline, \emph{ClaimCompare}, designed to generate labeled patent claim datasets suitable for training IR and ML models to address this challenge of novelty destruction assessment. To the best of our knowledge, ClaimCompare is the first pipeline that can generate multiple novelty destroying patent datasets. To illustrate the practical relevance of this pipeline, we utilize it to construct a sample dataset comprising of over 27K patents in the electrochemical domain: 1,045 base patents from USPTO, each associated with 25 related patents labeled according to their novelty destruction towards the base patent. Subsequently, we conduct preliminary experiments showcasing the efficacy of this dataset in fine-tuning transformer models to identify novelty destroying patents, demonstrating 29.2\% and 32.7\% absolute improvement in MRR and P@1, respectively.
\end{abstract}


\maketitle

\section{Introduction}
        Patent search is a rich and challenging space which comprises a diverse set of tasks, including Freedom to Operate (FTO) searches, novelty or patentability searches, and validity searches. Within this spectrum, patentability searches hold particular significance as they help gauge whether an invention's claims (i.e. structural features) are novel and non-obvious, and are therefore a critical part of the patent examination process. This assessment typically involves patent examiners, as well as inventors or their legal representatives, meticulously combing through prior art databases to uncover any existing disclosures that could potentially anticipate the invention and, consequently, undermine its novelty. In the United States, in particular, prior art is deemed ``novelty destroying'' if it anticipates or references every element of at least one of a proposed invention's claims.

        Traditionally, prior art searches have been performed manually, with searchers iteratively crafting and revising complex keyword and Boolean queries to obtain the most relevant documents. However, with the number of patent applications and volume of prior art growing annually, the labor-intensive nature of this process has become increasingly unsustainable, driving a growing interest in the usage of information retrieval (IR), machine learning (ML), and deep learning (DL) approaches to streamline search methodologies and optimize result relevance, for example, via query expansion and targeted semantic similarity techniques \cite{Aras2018GetYH, Helmers2019Auto, Navrozidis2020Using}. These advances build on prior work in the patent space pertaining to automated patent landscaping and other automation tasks related to patent code classification and categorization \cite{Abood2018Auto, Choi2019Deep, Grawe2017Auto, Fall2003Auto}. Far less work has focused on finding novelty destroying prior art; in fact, to date, there is only one other public dataset dedicated to this task \cite{Risch2020Patentmatch}, and none on US Patent data.
    
    \textbf{Contributions and Scope.} In this paper, we introduce \emph{ClaimCompare}, a novel data pipeline for generating patent claim datasets, labeled with respect to the novelty destruction search problem, in order to facilitate improved performance in this space. To the best of our knowledge, ClaimCompare is the first pipeline that can generate multiple novelty destroying patent datasets. We leverage publicly available United States Patent and Trademark Office (USPTO) APIs in order to curate such datasets, alongside web-scraping Google Patents. To simplify the problem, we focus only on identifying novelty destroying patents, rather than all potential literature contributing to novelty destruction. Our contributions are as follows: 

        \begin{itemize}[leftmargin=*]
            \item We construct \emph{ClaimCompare}, a pipeline utilizing the USPTO API and Google Patents to generate curated novelty destroying datasets.
            \item We utilize ClaimCompare to curate a sample dataset of 27K patents in a specialized domain, comprising of 1,045 base patents and 25 related patents for each. Of the base patents, 357 (34\%) have one or more identified novelty destroying patent(s). 
            \item To assess the effectiveness of ClaimCompare and the sample dataset, we perform experiments utilizing LLMs fine-tuned on our dataset to assist with novelty determination. Our experiments demonstrate 29.2\% and 32.7\% absolute improvement in MRR and P@1, respectively, over a baseline model. 
        \end{itemize}
    
        We envision ClaimCompare being used to generate both generic and domain-specific training datasets at scale, focused specifically on the task of novelty determination. These datasets can subsequently be used to train and test a variety of IR, AI/ML, and/or DL models for this task. We release all our pipeline code and data\footnote{\href{https://github.com/RIET-lab/claim-compare}{https://github.com/RIET-lab/claim-compare}}.

\section{Prior Work}
    \begin{table*}
        \centering

        \begin{tabular}{lccccccc}
        Dataset & Size & Data Source & Positive Samples & Negative Samples & Matching Strategy & Balanced? \\\hline
        PatentMatch v2 & 25K & EPO & Search report ``X" citations & Search report ``A" citations & Specific excerpts/lines & Yes\\
        CC Sample & 27K & USPTO & Office action 102 rejections & Similar keyword patents & Entire claim sets & No
        \end{tabular}

        \setlength{\abovecaptionskip}{-10pt}
        \caption{\label{tab:dataset_comp} PatentMatch vs. a sample ClaimCompare (CC) dataset. 
        \textmd{Note that the ClaimCompare pipeline can be used as-is in order to generate many other datasets, including significantly larger ones.}}
    \end{table*}
    
    Despite a very rich literature on patent search overall \cite{Krestel2021Survey}, there is little work on the important, highly-specialized task of detecting novelty destroying patents. 

    The now discontinued CLEF-IP tracks in 2012 and 2013 present useful datasets related to patentability searching. The 2012 edition released a claims to passage dataset containing 2.3 million European patent documents with  2,700 corresponding relevance judgements \cite{Gobeill2012Bitem}. The shared task is to create the most effective passage retrieval system given a particular topic. However, the task focused on claim sets, rather than single claims, and considers ``X” (novelty destroying) and ``Y” (inventiveness destroying) passages as equally relevant, merging the two problems rather than isolating novelty destruction. 
    
    A key paper on the task of novelty evaluation is PatentMatch \cite{Risch2020Patentmatch}, which offers the first dataset directly addressing novelty destruction in patents by leveraging European Patent Office (EPO) search reports. The dataset is composed of pairs, each containing an individual patent application claim paired with a passage from either an ``X" citation or an ``A" (background) citation. Unfortunately, as the authors note, fine-tuning a BERT model on the dataset produces relatively poor results with an accuracy of 54\% in the best case. A follow-up empirical study utilizing the dataset also found fine-tuned BERT and SBERT models to perform poorly with accuracies of 54\% and 57\% respectively \cite{Chikkamath2020Empirical}. In examining the PatentMatch dataset further, we see that many of the excerpts are quite short in length, lacking the context of the broader patent, which helps explain why context-dependent transformer models like BERT struggle to effectively capture the nuanced semantic relationships defining novelty destruction between patents. In light of this observation, we opt against incorporating the specific excerpts into our dataset, favoring instead the inclusion of broader claim sets which concisely encapsulate the essential elements that define the novelty of an invention, or lack thereof, while still providing sufficient context. 
    
    To the best of our knowledge, our paper is the first to approach novelty destruction from a US-centric approach, and offers the first public dataset for this task as well as a pipeline to easily generate additional datasets\footnotemark[1].

\section{Methodology}  
\begin{figure*}[htb]
            \centering
            \includegraphics[width=0.9\textwidth]{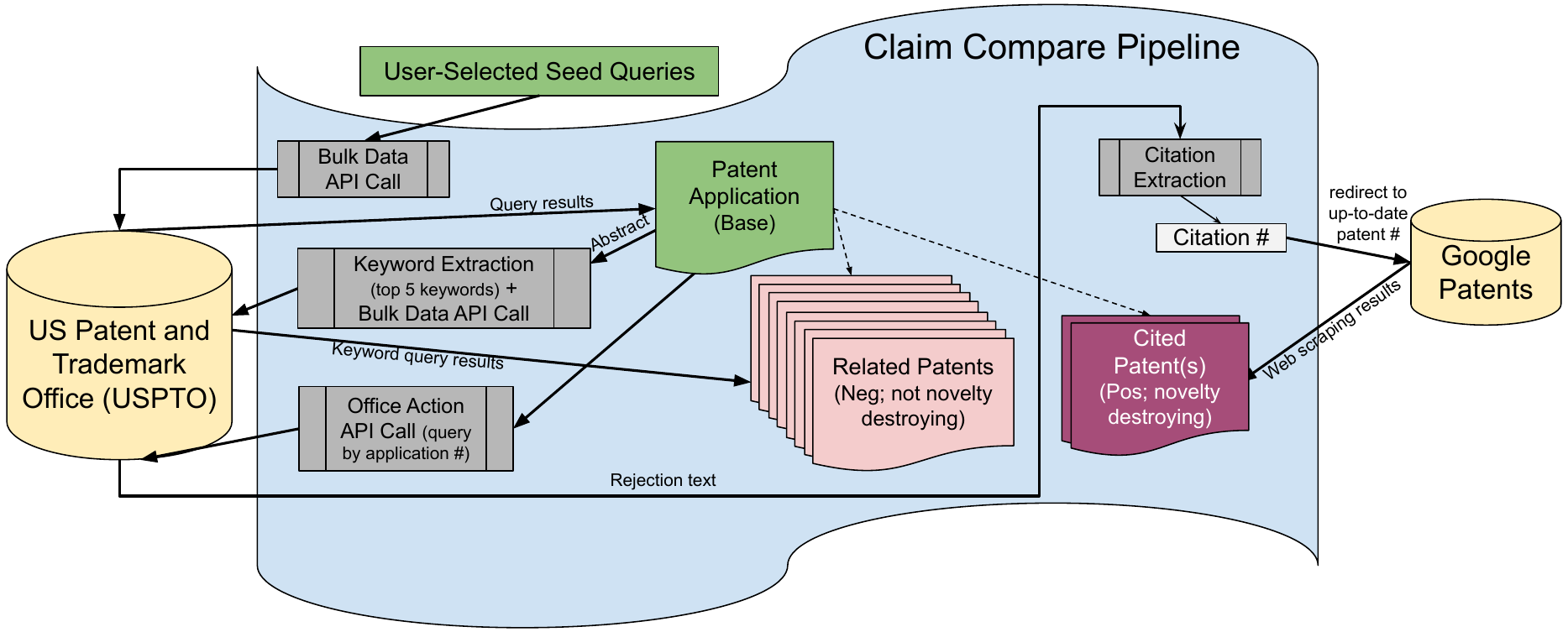}
            \setlength{\belowcaptionskip}{-12pt}
            \caption{\label{fig:pipeline} The ClaimCompare Pipeline. \textmd{ClaimCompare can accept any initial seed queries and generate a novelty destroying dataset for that query set. For each base patent, the pipeline finds $k$ novelty destroying and related, non-novelty destroying patents. The pipeline utilizes two types of API calls to the USPTO (Bulk Data and Office Action APIs) and scrapes data from Google Patents to account for patent number changes.}}
        \end{figure*}

    We now introduce the ClaimCompare data pipeline (Figure \ref{fig:pipeline}) which generates domain-centric and agnostic datasets in order to train DL models to assess patent novelty. For the remainder of the paper, we focus on describing our pipeline, sharing information about the sample dataset, and demonstrating its effectiveness for this task via preliminary experiments involving large language models (LLMs) fine-tuned on this sample dataset.
    
    \subsection{Approach}
        To ground the use of our ClaimCompare pipeline in the context of novelty determination specifically, we define the novelty destruction search problem as a sub-problem of the prior art retrieval process. Accordingly, ClaimCompare operates under the assumption that a preliminary set of prior art patents of size $k$ has already been retrieved by the searcher for a given query (i.e., base patent $q$), using preexisting methods. While the models trained with ClaimCompare's datasets can certainly be used to improve direct retrieval of only novelty destroying patents (i.e., when $k$ is made sufficiently large), in this paper we focus on applying these models as a filter on top of a smaller subset of retrieved patents.
        
        To develop ClaimCompare, we primarily rely on two publicly available USPTO APIs to access the patent data necessary to teach subsequently trained models to semantically differentiate between novelty destruction and mere relevance. Relevant patents constitute ``negative" samples in our dataset and are queried for with keywords extracted from the base patents, mimicking a common prior art search practice. In contrast, novelty destroying ``positive" samples are generated by obtaining USPTO office actions with rejections citing novelty destroying prior art patents. For our positive samples, we only consider office actions that contain a 102 rejection, indicating that the citation is considered to be novelty destroying for the corresponding application by skilled patent examiners. Our focus on 102 rejections is deliberate, as they only reference elements from a single document as sufficient to destroy the novelty of a patent application, whereas the more nuanced and complex \textit{103 rejections} often refer to elements from multiple patents and/or common technical knowledge coming together to invalidate the novelty of the proposed patent.

    \subsection{Implementation Details}
        ClaimCompare's pipeline starts with a set of seed queries, which are sent to the USPTO Bulk Data API\footnote{\href{https://developer.uspto.gov/api-catalog/bulk-search-and-download}{https://developer.uspto.gov/api-catalog/bulk-search-and-download}}. In the context of the provided sample dataset, we use the phrase ``redox flow battery" as a query to retrieve inventions in the electrochemical device space; naturally, this can be replaced with keywords/phrases in any given domain.
        
        For each retrieved patent application, we collect the application and publication numbers, abstract, and claims in order to form our set of base patents; we then set out to acquire their cited novelty destroying patent(s), if applicable, and other related patents. For the cited patents, we query the USPTO Office Action Citation API\footnote{\href{https://developer.uspto.gov/api-catalog/uspto-office-action-citations-api-beta}{https://developer.uspto.gov/api-catalog/uspto-office-action-citations-api-beta}} using the base patents' application numbers. If an office action is found with a 102 rejection, we take the rejection text and pass it to a text2text generation base T5 model along with a standardized prompt in order to extract the publication number of the novelty destroying patent cited within the text. The T5 performs the task efficiently with a 94\% success rate, which is sufficiently high for our needs. We then perform a simple cleanup on the publication number and use it to query for its claims via Google Patents\footnote{\href{https://patents.google.com}{https://patents.google.com}}, which can web redirect to the most up-to-date version of the patent if the number is outdated, unlike the APIs. To acquire our negative samples, we extract keywords and phrases from the base patent abstract. We apply the KeyBERT model to get the top 5 keywords from the abstract \cite{Grootendorst2020Keybert}, with which we query the USPTO Bulk Data API for the number of relevant patents it takes to meet the limit $k$ per base patent depending on the number of positive samples previously acquired. We omit smaller details of the related patent query process for space considerations; we refer interested readers to our codebase. 
        
    \subsection{Dataset Structure}
        Of the 1,045 rows in our raw sample dataset, 357 (34\%) of them contain at least one positive sample. Of these 357 rows, 36 (10\%) have two positive samples; there are no rows with three or more novelty destroying patents in our dataset. To ensure that the number of samples per row is always $k$, we set $Claims\_25$ and/or $Claims\_24$ and their corresponding publication numbers as null values, depending on the number of novelty destroying patents present. If none are found for a given base patent, then the rejection columns are set as null instead. 
        
        Clearly, given the structure of our dataset, there is an inherent lack of balance between the two classes. We deliberately maintain this imbalance for two main reasons. Firstly, it reflects the current state of prior art searching where there are far more relevant samples than novelty destroying samples for any given patent. While imbalanced datasets typically pose challenges for training ML or DL models, we are intrigued to explore how this realistic representation of class distribution influences model performance in our experimentation. Secondly, although we do not directly train a ranking model in our experimentation, we indirectly test the ability of our model to rank prior art patents based on their likelihood of invalidating the novelty of a given base patent, a task which necessitates a larger value for $k$ and, consequently, an imbalanced dataset.

\section{Experimental Setup and Results}
    In order to assess the effectiveness of our dataset in training LLMs to assist with novelty determination, we test whether fine-tuning these models outperforms a baseline pre-trained BERT-based transformer model.
    
    \subsection{Training Data}
        To prepare our raw sample dataset for model training, we drop all non-claim columns and convert the row-wise format of the dataset into a pairwise format such that each base patent is individually matched with each of its relevant or novelty destroying patents to form the training examples. In other words, instead of having rows where a base patent is matched with 25 related patents, there are now 25 rows enumerating each of these matches. The base patents are found in the $Claims\_x$ column while their related matches are each found in the $Claims\_y$ column. If the pair is novelty destroying, the $Label$ column contains a 1 to denote the positive match; otherwise, it contains a 0.

        We use an 80-10-10 stratified train-val-test split, with each of the splits possessing roughly the same proportion of positive samples. To avoid data leakage, the base patents are restricted to one of the splits such that all 25 pairs can be found in that split. To mitigate the effects of sampling bias, we randomly subsample the raw dataset twice more to generate a total of three unique train-val-test splits for our model. The results we present are all averaged across these three runs. 

        We also intentionally downsample the negative samples in our training dataset, to observe the effect this has on both the validation and testing metrics. We perform this downsampling by simply reducing $k = 25$ to $k = 10$ and $k = 5$ on our three shuffled train-val-test splits, retaining all positive sample while randomly sampling the required number of negatives examples from the larger set for each base patent.

    \subsection{Model Fine-Tuning}
        For our experiments, we fine-tune a sequence classification model with our sample dataset. We primarily use the base DistilRoBERTa model \footnote{\href{https://huggingface.co/distilbert/distilroberta-base}{https://huggingface.co/distilbert/distilroberta-base}} due to its compact size and robust performance on related tasks \cite{Sanh2019Distilbert}. We also use the BERT for Patents model \footnote{\href{https://huggingface.co/anferico/bert-for-patents}{https://huggingface.co/anferico/bert-for-patents}} as a stronger, domain-specific baseline but unfortunately lack the computational resources to fine-tune such a large model. As a result, we choose to fine-tune the DistilRoBERTa model ``from scratch," presenting an intriguing opportunity to assess the ability of this model to adapt to both the broader patent domain and our specific novelty determination use case. We train the model for 3 epochs with a cross entropy loss function, training and validation batch size of 16, learning rate of 0.00002, and weight decay of 0.01.

        \begin{table}
            \centering

            \begin{tabular}{lcccc}
            Model & AUROC & AP & MRR & P@1 \\\hline
            General Baseline & 0.473 & 0.350 & 0.697 & 0.651\\
            Domain Baseline & 0.589 & 0.464 & 0.703 & 0.651\\
            Fine-Tuned ($k = 25$) & 0.999 & 0.999 & 0.989 & 0.978\\
            Fine-Tuned ($k = 10$) & 0.999 & 0.998 & 0.987 & 0.975\\
            Fine-Tuned ($k = 5$) & 0.982 & 0.975 & 0.967 & 0.934\\
            \end{tabular}
            
            \setlength{\abovecaptionskip}{-16pt}
            \caption{\label{tab:test_res} Model testing results.
            \textmd{General baseline and fine-tuned models rely on DistilRoBERTa. Domain baseline relies on BERT for Patents.}}
        \end{table}
      
    \subsection{Model Evaluation and Discussion}
        Once the model has been trained, we assess its performance on our testing dataset. However, rather than solely test its ability to perform pairwise classifications, we take each set of 25 test patents and combine the model's pairwise predictions for each of the patents in the set using a simple logical OR as an ensemble, such that if any patent in the set is deemed to be novelty destroying, the base patent is found not novel. To quantify the model's performance on this task, we compute classification metrics suitable for our imbalanced dataset such as average precision (AP) and area under the receiver operating characteristic curve (AUROC). Since these metrics require prediction scores as opposed to labels to compute, we take the maximum novelty destruction score (i.e., pairwise probability of the positive class) out of all the patents as the representative score for the set. 
        
        Additionally, we also evaluate the model in the context of ranking, using their novelty destruction scores for rank ordering the test patents in each set of 25. We use ranking metrics such as MRR and Precision@1 to assess the performance. The unique nature of the novelty destruction task means there are often no members of the positive class in a given set; we therefore introduce a placeholder patent, assigned a score of 0.9, in order to accurately compute these metrics. The placeholder patent is inserted into the set prior to ranking, in order to denote a patent which the model regards as reasonably novelty destroying. To facilitate accurate comparisons across the board on the testing set, we introduce the placeholder patent to all of the patent sets, such that if a novelty destroying patent is actually present, it is ideally ranked first, above the placeholder patent. Alternatively, when no novelty destroying patent is present, the placeholder patent is ideally ranked \#1. 
        
        Examining the impact of fine-tuning with our dataset on model performances (Table \ref{tab:test_res}), we see significant improvement over both baselines. As expected, the models perform quite poorly in the baseline, zero-shot setting, especially in terms of classification, with BERT for Patents exhibiting only a slight improvement over DistilRoBERTa. The near-perfect classification and ranking performance metrics we see after fine-tuning, however, are quite surprising. The high AUROC and AP, in particular, indicate that the model is able to differentiate between the positive and negative samples with a high degree of confidence. The $k = 25$ model is the top performer, but with little distinguishing it from the $k = 10$ model. This is intriguing as the model appears to be performing better in spite of the greater imbalance, perhaps simply due to the presence of more data allowing the model to better learn the relationships defining novelty destruction. 
        
    \subsection{Limitations and Future Work}
        These results, though incredibly promising, point to some limitations due to the large gap between the baseline and fine-tuned models, and exceptionally high testing scores. While we can confidently say that the fine-tuned models are more suited for the novelty evaluation task than the base models, further comparisons are needed to baseline models pre-trained on general legal data, which would likely provide a stronger baseline; however, due to computational resource limitations at submission time, we leave this to future work. 
        
        We note that our models do not appear to be overfitting, given that the loss trends during training are nominal, and these models are generalizing well to the unseen data of both the validation and testing sets. Manual examination of a significant portion of the dataset shows no sign of data leakage, and the training appears to be sound as well. Thus, we hypothesize that the high absolute results are the consequence of the negative samples selected via keyword search being far too ``easy'' to differentiate from the positive, novelty destroying samples. Whether this issue stems from the keyword extraction process itself, the fact that more powerful Boolean queries are needed to obtain the most relevant results, or perhaps even another unseen factor, requires further experimentation. We leave an in-depth exploration of how to leverage unique inter-patent relationships to build upon our pipeline to future work; likewise with the task of finding more semantically similar non-novelty destroying patents (i.e., harder negatives) to match with the application to improve robustness at the classification boundary. We note that the USPTO has a new Citation API that can potentially be of use for these goals.

        We also invite future work to employ ClaimCompare in order to generate additional datasets, inclusive of additional technical fields, for example, by creating queries that utilize Cooperative Patent Classification (CPC) codes instead of keywords. With these datasets, future work can train state-of-the-art models, such as generative LLMs, to improve performance even further.
    
\section{Conclusion}  
    In this paper, we introduce a novel pipeline, \emph{ClaimCompare}, to generate datasets geared towards patent novelty evaluation; offer a sample dataset generated using ClaimCompare; and assess its utility in fine-tuning a novelty destroying classifier. We leverage USPTO APIs to obtain our novelty and non-novelty destroying data. To the best of our knowledge, this is the first usage of USPTO office actions and patent claims as a source of data, providing high quality datasets while being fairly straightforward, flexible, and easy to replicate.

    We believe ClaimCompare holds potential to accelerate research in the patent retrieval field, in conjunction with the use of LLMs and other cutting-edge DL models. Improving novelty determination holds the promise of reducing the time and monetary burden required for patent search and patentability determinations, while also increasing accuracy, thereby saving searchers' time and making patent databases more accessible for all. In this sense, we hope that our pipeline can facilitate the democratization of what has been a historically complex process, and enable inventors, attorneys, and patent examiners alike to assess patent novelty with greater ease.

\small\bibliography{refs}

\end{document}